\documentclass[aps,prl,twocolumn]{revtex4-1}
\usepackage{graphicx}% Include figure files
\usepackage{dcolumn}% Include figure files

\begin{document}

%\draft

\title{Localization and shock waves in curved manifolds for the Gross-Pitaevskii equation}
%{\large{\bf }}
\author{Claudio Conti$^{1}$}
\affiliation{
$^1$Department of Physics, University Sapienza, Piazzale Aldo Moro 2, 00185, Rome (IT)}
\email{claudio.conti@uniroma1.it}
\date{\today}

\begin{abstract}
We investigate the dynamics of a Bose-Einstein condensate in a progressively bended 
three dimensional cigar shaped potential. The interplay between geometry and nonlinearity is considered. At high curvature, topological localization occurs and becomes frustrated
by the generation of curved dispersive shock-waves when the strength of nonlinearity is increased.
The analysis is supported by four-dimensional parallel simulations.
\end{abstract}

%\pacs{}

\maketitle
The effect of geometry on wave propagation has been considered since early developments of the theory of sound, 
for example, in the vibrations of membranes with different shapes \cite{RayleighBook1877}.
Geometry largely affects energy velocity, wave transformations upon propagation, and various classical and quantum undulatory phenomena, 
either in constrained geometries, or at an astrophysical scale; renowned examples are the Einstein lensing effect\cite{Einstein1936}, the Hawking radiation \cite{HAWKING1974},or the Unruh effect \cite{unruh1976}

Albeit this old history, the link between geometry and waves is continuously fascinating many researchers, and is currently driving important applications as transformation optics \cite{Longhi2007,Leonhardt2008}, curved and twisted waveguides, \cite{Longhi2007,Wong2012} and analog of gravity in  Bose-Einstein condensation (BEC) \cite{Barcelo2003,pavloff2012} or nonlinear optics 
\cite{Skryabin2007,peschel2010,faccio2012}.
It is also known that nonlinear waves may be affected by geometrical constraints, resulting in interesting phenomena, as solitons in curved manifolds \cite{batzpeschel2008,batzpeschel2010}.

A striking effect of curvature is the topologically induced localization, originally considered in \cite{daCosta1981}: when the wave is constrained on extremely deformed surfaces, or lines, 
propagation may be inhibited and energy is trapped in the regions with the highest curvature. The way this kind of geometrical localization competes with nonlinearity, and the kind of dynamic effects resulting from a nonlinear response 
sufficiently strong to overcome topological bounds is largely unconsidered.

If geometrical localization occurs in a reduced dimensionality (e.g., a curved surface in a three dimensional space), when nonlinearity is very effective the whole three-dimensional (3D) space becomes involved, and any treatment based on nonlinear wave equations with reduced dimensionality may be questioned. This is a key difficulty in this problem; any theoretical prediction must be numerically tested by using 3D simulations.

In addition, for high nonlinearity, shock waves (SW) are generated \cite{WhithamBook} . In recent years, SW have been largely considered in optics and BEC \cite{Damski2004,Gentilini:12,Hoefer2006,EL2007,Wan2007}; 
SW originate from singular solutions of the hydrodynamic reduction of the nonlinear Schroedinger equation (NLS) regularized by oscillating wave fronts, named undular bores, or dispersive SW (D-SW). To the best of our knowledge, the effect of a curved space on DSW is un-explored.

In this Letter, we investigate the way a geometrical localization is frustrated by a defocusing nonlinearity, we study this effect by one-dimensional (1D) theoretical analysis and 3D+1 simulations of the Gross-Pitaevskii (GP) equation,
and report on DSW in curved potentials.
\begin{figure}
\includegraphics[width=0.48\textwidth]{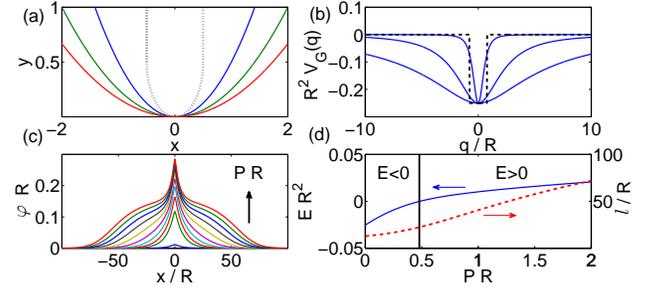}
\caption{(Color online) 
(a) Potential profile for the piecewise case
(dashed, $R=0.5$) and for the parabolic cases ($y=k x^2$ with $R=1/2k=1,2,3$);
(b) geometrical potential $V_G$ for the cases in panel (a);
(c) shape of the bound state for the parabolic potential for various powers 
($P R$ goes from $10^{-3}$ to $2.6$); (d) nonlinear eigenvalue versus power (left axis), and localization length (right axis), 
$w_1 R^2=6.25\times 10^{-6}$. 
\label{fig1} }
\end{figure}

\noindent {\it The model ---}
The adimensionalized GP equation is
\begin{equation}
i \Psi_t=-\nabla^2 \Psi+V_{3D}(r)\Psi-\chi |\Psi|^2 \Psi\text{.}
\label{GP3D}
\end{equation}
In (\ref{GP3D}), $\chi=\pm 1$ determines the sign of the nonlinearity, and $N=\int |\Psi|^2 d^3 \mathbf{r}$.
%$N=\int |\Psi|^2 d^3 \mathbf{r}$.
%\begin{equation}
%N=\int |\Psi|^2 d^3 \mathbf{r}\text{.}
%\end{equation}
We consider a curved potential, sketched in figure \ref{fig1}, $V_{3D}=V_1(q_1)+V_\perp(q_2,q_3)$ where $q_1=q$ is the {\it longitudinal} coordinate along the arc, and $q_{2,3}$ 
are the {\it transverse} coordinates.\cite{daCosta1981} 
$V_\perp$ determines the transverse confinement along an arbitrary curve,
whose curvilinear coordinate is given by $q$.
$V_1(q_1)$ is a weak longitudinal trapping potential, whose effect becomes negligible for strong curvature, as detailed below. 
Specifically, we have a parabolic potential $V_\perp=w (q_2^2+q_3^2)$, $V_1(q_1)=w_1 q_1^2$ with $w>>w_1$. 
Being $\lambda^2=w/w_1$, the 1D reduction holds true as 
$\lambda^2\rightarrow \infty$ \cite{daCosta1981}, letting
\begin{equation}
\Psi(q_1,q_2,q_3)=l_\perp \psi_\perp(q_2,q_3) \psi(q_1)\exp\left(-i E_\perp t\right)
\end{equation}
with $\int |\psi_\perp|^2(q_2,q_3) dq_2 dq_3= 1$, and $E_T$ the 
trasverse part of the eigenvalue.
\begin{figure*}
\includegraphics[width=\textwidth]{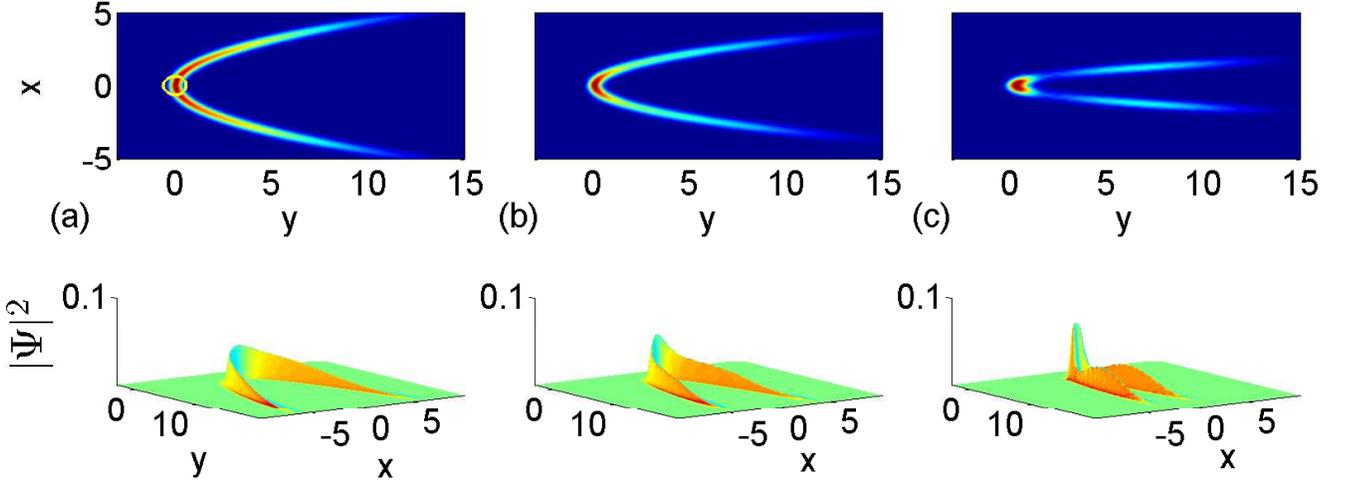}
\caption{(Color online) 
Snapshots of the two dimensional section of the wave-function 
in the bending plane $xy$ 
for different curvatures in the linear case
($\chi=0$, $t=1.5$); (a) $k=0.5$, (b) $k=1$ and (c) $k=4$
}
\label{fig_diffusion} 
\end{figure*}
%
%\begin{equation}
%\int |\psi_\perp|^2(q_2,q_3) dq_2 dq_3= 1
%\end{equation}
The transverse localization length is 
$l_T^{-2}=\int |\psi_\perp(q_2,q_3)|^4 dq_2 dq_3$,
%\begin{equation}
%l_T=\left(\int |\psi_\perp|^4 (q_2,q_3) dq_2 dq_3\right)^{-1/2}
%\end{equation}
and the normalization condition $\int |\psi(q)|^2 dq=N l_\perp^{-2}\equiv P$.
%\begin{equation}
%\int |\psi(q)|^2 dq=\frac{N}{l_\perp^2}\equiv P\text{.}
%\end{equation}
For the parabolic potential above, one has for the linear ground state $\psi_\perp=\sqrt{2/\pi l_T^2} \exp(-q_2^2/l_T^2-q_3^2/l_T^2)$ with $l_T=(4/w)^{1/4}$, and $E_T=2 \sqrt{w}$. The resulting 1D GP equation with geometrical potential $V_G$ is given by
\begin{equation}
i \partial_t \psi=-\partial_{q}^2 \psi+V(q)\psi- \chi |\psi|^2 \psi
\label{NLSreduced}
\end{equation}
where $V=V_1+V_G$, and $V_G$ is expressed in terms of the local curvature $K(q)$:
\begin{equation}
V_{G}(q)=-\frac{ K^2(q)}{4}\text{.}
\end{equation}
The bound state can be determined by the normalized one-dimensional equation 
\begin{equation}
-\varphi_{qq}+V(q)\varphi -\chi \varphi^3=E\varphi(q)\text{.}
\label{bound1d}
\end{equation}
\\\noindent {\it Linear geometrical localization ---}
For $\chi=0$ and $V_1=0$, we consider a step-wise curvature \cite{daCosta1981},  as in figure \ref{fig1}a, dashed line: $V_G(q)=0$ for $|q|>R$ and $V_G(q)=-1/(4R^2)$ for $|q|<R$ with $R$ the minimum radius of curvature. This piecewise potential allows simple analytical results: there is a single bound state, which is exponentially localized with respect to $q$ and exponent $\sqrt{-E}$, with $E\simeq -0.1/R^2 <0$. 

\noindent For a bended cigar shaped trap (Fig.\ref{fig1}a), with profile $y=k x^2$,  we have
\begin{equation}
V_G(q)=-\frac{K(q)^2}{4}=-\frac{k^2 }{\left[1+4 k^2 x(q)^2\right]^3}
\label{vgeo1}
\end{equation}
where $x(q)$ is given by the inverse of $4 k q=2 k x \sqrt{1+4 k^2 x^2}+\sinh^{-1}(2 k x)$. The maximum curvature is $2k=1/R$. 
As for the stepwise potential, 
a unique eigenstate with negative energy exists with $E\cong -0.1/R^2$, as numerically calculated. An analytic approximation is found by replacing $V_G$ by a Dirac $\delta$ with the same area, this furnishes $E\cong -1/(9R^2)$ and decaying tail $\propto\exp(-|q|/l_G)$, with localization length  $l_G\cong 4 R$ of the order of $R$. 
\\When also considering the presence of a longitudinal trapping potential $V_1(q)=w_1 q^2$, the associated Gaussian bound state $\varphi=\sqrt{\sqrt{2/\pi} P/l_1} \exp(-q^2/l_1^2)$ with $l_1=\sqrt{2}/w^{1/4}$,
may mask the exponential localization as far as 
$l_1$ is comparable with the local radius of curvature; 
geometric localization appears for $l_G<<l_1$.
\\\noindent {\it Nonlinearity and curvature ---}
For the piecewise curvature, in the presence of defocusing nonlinearity ($\chi=-1$), the Thomas-Fermi approximation allows to write $\varphi^2=E-1/(4 R^2)$ for $|q|<\pi R/2$ and $\varphi=0$ elsewhere \cite{Dalfovo99}. The normalization $\int \varphi^2 dx=P$ provides the power dependent nonlinear eigenvalue $E=-1/(4 R^2)+P/(\pi R)$, which shows that, for $P=P_{th}=\pi/(4R)$,
$E$ changes sign. For $P>P_{th}$, $E>0$, and the wave is delocalized: the  nonlinearity destroys the geometrical localization.

For the parabolically bended cigar-shaped potential for $\chi=-1$, we numerically solve the nonlinear bound-state equation (\ref{bound1d}) by a pseudo-spectral Newton-Raphson approach, also including the longitudinal potential $V_1=w_1 q^2$. In figure \ref{fig1}c we show the calculated profiles for different powers; 
the results can be expressed in terms of variables suitably scaled by $R$ without loss of generality. When increasing $P$, there is a transition from an exponentially localized state to the Gaussian ground state of the longitudinal potential; correspondingly $E$ changes sign, from negative to positive (Fig.\ref{fig1}d), and the localization length increases. When $V_1=0$ a transition to a delocalized state is found but not reported.
\begin{figure}
\includegraphics[width=0.45\textwidth]{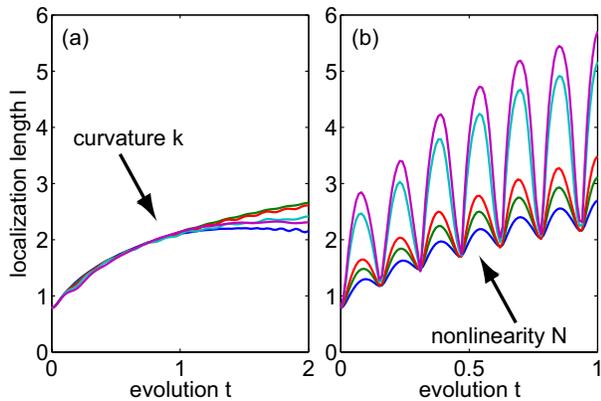}
\caption{(Color online) (a) Linear case ($\chi=0$), localization length versus time for various values of the curvature $k$ ($k=0,0.5,1,2,4$);
(b) nonlinear case ($k=1$, $\chi=-1$), localization length versus time $t$ for various number of atoms ($N=10,20,30,100,500$).
The localization length is calculated as the three-dimensional inverse participation ratio.
\label{fig_loclength} }\end{figure}
Note that the power for delocalization scales as the inverse of the minimum radius of curvature, i.e., the more the curvature, the higher the strength of nonlinearity needed to delocalize.
This in agreement with the simple theoretical calculations for the stepwise potential reported above. We numerically find $P_{th}\cong 0.5/R=k$.

\noindent{\it Numerical results in the linear case ---}
We solve the 3D+1 GP equation 
by a pseudo-spectral parallel algorithm. 
We first consider the case without nonlinearity, and show the way the geometrically induced localization appears in the spreading of an initially
localized wave packet in a cigar shaped potential when increasing curvature.
We consider an initial wavefunction that matches the transverse size of the linear ground state for the transverse potential $V_\perp = w q_\perp^2$ with $w=100$;
the initial condition is $\Psi= \exp(-q^2/l_1^2)\exp(-q_2^2/l_T^2-q_3^2/l_T^2)$ 
with $l_1=l_T=(4/w)^{1/4}=0.44$, other parameters are $\lambda^2=10^5$ and $V_1=w_1 q_1^2$ and $w_1=10^{-3}$, 
the initial size is indicated by the circle in fig.\ref{fig_diffusion}a.

When starting from low bending ($k<1$), as the initial longitudinal length is much smaller ($l_1=0.44$ in our units) than the size of the fundamental state ($l_1=8$), 
the wave spreads in the $q$ direction,  as shown in figure \ref{fig_diffusion}a.
When the curvature increases, the wave displays a degree of localization, corresponding to the excitation of the topologically trapped state, as shown in Fig.\ref{fig_diffusion}b,c. 
The related localization length is of the order of the radius of curvature.
Fig.\ref{fig_loclength}a shows the localization length calculated by the inverse participation ratio for an increasing curvature; at high curvature the localization length asymptotically becomes nearly time-indepedent.
%%%%%%%%%%% SHOCK IN THE NONLINEAR CASE

\noindent {\it Hydrodynamic approximation ---}
In the presence of a defocusing nonlinearity, the geometrical localization is opposed to the tendency to spread.
Following the analysis above, one expects a power threshold for the nonlinear frustration of the topological trapping; we show in the following that this is mediated by the generation of DSW.

We consider an initial wave-packet with Gaussian shape and waist
of the order of the curvature, as in Fig.\ref{fig_diffusion}.
We investigate the expansion of this condensate in the presence
of a defocusing nonlinearity and of the geometrical potential.
The theoretical analysis can be developed in the hydrodynamical approximation
\cite{Ghofraniha07}. Within the 1D model, we consider the evolution of an Gaussian beam $\psi(t=0)=\exp(-\eta^2/2)$ with $P=\sqrt{\pi}/\epsilon$. 
Introducing $\tau=t/\epsilon$, $\eta=\epsilon q/ \sqrt{2}$, the scaled equation reads as
\begin{equation}
i\epsilon\psi_{\tau}=-\frac{\epsilon^2}{2} \psi_{\eta\eta}+V_G \psi+|\psi|^2 \psi=0
\end{equation}
In the limit $\epsilon\rightarrow 0$, the 
occurrence of an hydrodynamic shock is signaled by collapsing trajectories
$\eta(\tau)$ solutions of the equation \cite{Ghofraniha2012}
\begin{equation}
\frac{d \eta^2}{d \tau^2}=-\frac{\partial V_{hydro}}{\partial \eta}=
-\frac{\partial}{\partial \eta}\left[V_G(\frac{\epsilon \eta}{\sqrt{2}})+exp(-\eta^2)\right].\end{equation}
The dynamics is determined by the potential
$V_{hydro}=V_G+V_\rho$, with $V_\rho\equiv \exp(-\eta^2)$ given by the initial Gaussian wave-profile. In the absence of curvature $V_G=0$ and the shock is determined by
an inverted well Gaussian potential $V_\rho$; 
in this case it is well known that the wave develops a couple of 
symmetrical lateral shocks  \cite{Ghofraniha2012}.
In the presence of curvature, as $V_G<0$ and $V_\rho>0$, the geometry seems to prevent the shock, however, due to the specific properties of the topological potential, this effect is negligible and curved DSW are expected.

Indeed, following Eq.(\ref{vgeo1}), the peak value of the geometric potential
is $k^2$ and its width in the scaled variable $\eta$ is $w_\eta\simeq \epsilon/k$,
while the width of $V_\rho$ is of the order of the unity.
There are hence two possibilities:
(i) when starting from a wavefunction with size larger than the
radius of curvature, the width of the potential well $w_{\eta}$  is much narrower than the initial wave profile, and $V_G$ does not
prevent the occurrence of the lateral shocks; 
(ii) when considering an initial condition with size comparable with the radius of curvature ($w_\eta\cong 1$, $\epsilon\cong k$), the maximum value of $V_G$, $k^2$, is much smaller than the unitary peak value of $V_\rho$
 and the shock occurs when $\epsilon<<1$, i.e., when $P>>\sqrt{\pi}$.

The geometrical potential and the related topological localization hence do not prevent the formation of DSW, which occur along the curved manifold. One may argue if the approximations underlying the reduced model Eq.(\ref{NLSreduced}) are still in the presence of strong nonlinearity,
this is confirmed in the simulations reported in the following.
\begin{figure*}
\includegraphics[width=\textwidth]{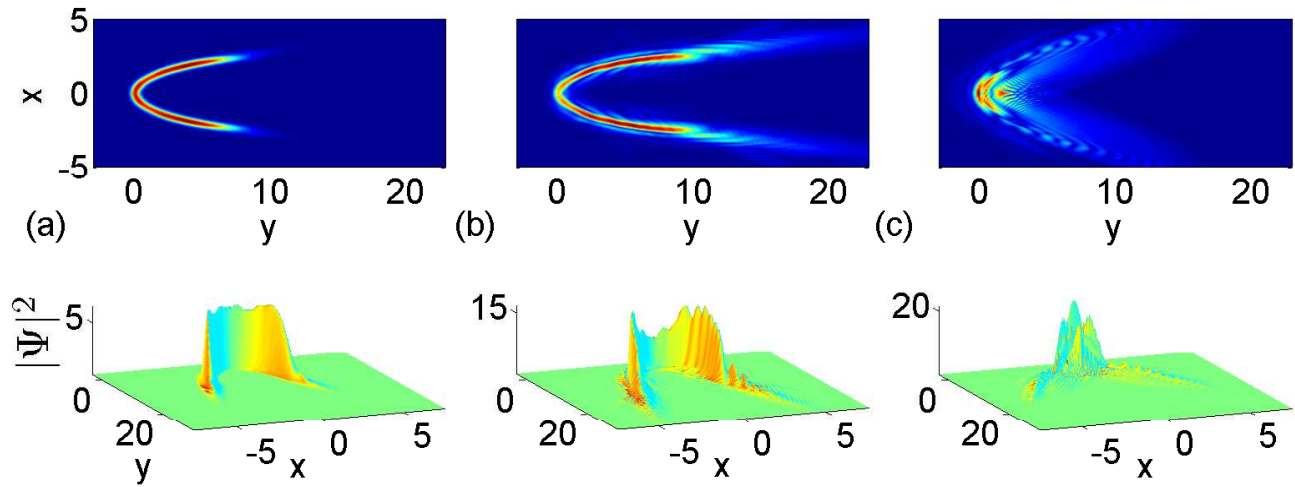}
\caption{(Color online) 
Snapshots of the nonlinear dynamics for an increasing number of atoms: 
(a) $N=25$, $P=130$, $t=0.6$; 
(b) $N=84$, $P=435$, $t=0.6$; (c) $N=168$, $P=870$, $t=0.2$.
Other parameters: $\chi=-1$, $w=100$, $w_1=10^{-3}$, $k=1$.}
\label{fig3d_nonlinear} \end{figure*}

\noindent {\it Dispersive shock waves in curved manifolds ---}
Here we consider the evolution of an initially localized wavefunction 
with nonlinearity; we study the $3+1$D evolution when increasing power. A representative series of snapshots are given in figure \ref{fig3d_nonlinear}, we show sections in the transversal plane $xy$ where curvature occurs, the wave stays confined by the trap in the $xz$ plane. The occurrence of the characteristic oscillation associated with the generation of the {\it undular bore} 
in the strong nonlinear regime is evident.
The shock formation is due an overwhelming nonlinearity 
that masks the geometrical trapping. The trend of the localization length versus time for increasing power is in figure \ref{fig_loclength}b,
which also shows oscillations typical of the undular bore.
\\{\it Conclusions ---}
We investigated the interplay between geometrically induced wave-localization
and nonlinearity for the Gross-Pitaevskii equation. 
For a defocusing nonlinearity, 
we determined the power needed for the transition to a delocalized regime
in terms of the curvature.
We have shown that at the frustration
of the geometrical trapping, the nonlinearity induces hydrodynamical
phenomena.
The theoretical analysis is supported by 
numerical simulations in four dimensions. 
Our results show the way complex manifolds in Bose-Einstein condensation,
and in related fields as nonlinear optics, may systain novel dynamical effects, as curved dispersive shock waves, and open the road to a variety of related analyses on the competition between geometry and wave-localization.

We acknowledge support from the ISCRA High Performance Computing initiative.
The research leading to these results has received funding from the European
Research Council under the European Community's 
Seventh Framework Program (FP7/2007-2013)/ERC grant agreement n.201766.

%%%%%%%%%%%%
%\bibliography{SSCMbib.bib}
%\bibliography{/home/claudio/incorso/bibtex/GIGAbib}
%\bibliography{../../../../../bibtex/GIGAbib}

\end{document}